\documentclass[journal=jacsat,manuscript=article]{achemso}

\usepackage[version=3]{mhchem} 

\usepackage{upgreek,amsmath} 
\usepackage{ulem} 
\usepackage{color}
\author{Binbin Wang}
\author{Yafei Ji}
\author{Linpeng Gu}
\author{Liang Fang}
\author{Xuetao Gan}
\email{xuetaogan@nwpu.edu.cn}
\affiliation[Northwestern Polytechnical University]
{Key Laboratory of Light Field Manipulation and Information Acquisition, Ministry of Industry and Information Technology, and Shaanxi Key Laboratory of Optical Information Technology, School of Physical Science and Technology, Northwestern Polytechnical University, 710129 Xi’an, China}
\author{Jianlin Zhao}

\title[SHG]
  {High-efficiency second-harmonic and sum-frequency generation in a silicon nitride microring integrated with few-layer GaSe
  }

\abbreviations{IR,NMR,UV}

\begin{document}







\begin{abstract}
  Silicon nitride (SiN) photonics platform has attributes of ultra-low linear and nonlinear propagation losses and CMOS-compatible fabrication process, promising large-scale multifunctional photonic circuits. However, the centrosymmetric nature of SiN inhibits second-order nonlinear optical responses in its photonics platform, which is desirable for developing efficient nonlinear active devices. Here, we demonstrate high-efficiency second-order nonlinear processes in SiN photonics platform by integrating a few-layer GaSe flake on a SiN microring resonator. With the pump of microwatts continuous-wave lasers, second-harmonic generation and sum-frequency generation with the conversion efficiencies of 849\%/W and 123\%/W, respectively, are achieved, which benefit from GaSe’s ultrahigh second-order nonlinear susceptibility, resonance enhanced GaSe-light interaction, and phase-matching condition satisfied by the mode engineering. Combining with the easy integration, the GaSe-assisted high-efficiency second-order nonlinear processes offer a new route to enriching already strong functionality of SiN photonics platform in nonlinear optics.
\end{abstract}
  

\section{Introduction}
Recently, silicon nitride (SiN) photonics is emerging in a growing number of photonic integrated circuit (PIC) applications. For instance, SiN-based waveguide components offer broader transparency spectral range from visible to mid-IR \cite{soref2010mid}, lower propagation loss \cite{lin2013low}, and better thermal stability \cite{arbabi2013measurements}. As a result, SiN photonics  presents superior properties for passive photonic devices and is becoming an ideal platform for microwave photonics \cite{liu2020photonic}, biophotonic sensing and imaging \cite{subramanian2015silicon}, quantum photonics \cite{taballione20198}, and high-power LiDAR (Light Detection and Ranging) \cite{poulton2017large}. More importantly, SiN is a flexible CMOS-compatible material, which enables its PIC platform with rapid developing progress and great potentials in commercialization assisted by the open-access foundry processes \cite{rahim2019open}.  

For a sophisticated PIC platform, it is desirable to have the ability of implementing nonlinear optical effects, which could greatly enrich PIC functionalities, including all-optically controlled devices \cite{pogna2021ultrafast}, light source generations with new spectral ranges \cite{photonics7040091}, entangled photon generations \cite{ramelow2015silicon}, etc. SiN PICs have been demonstrated as a decent platform to carry out third-order nonlinear optical effects relying on its ultra-low nonlinear optical loss and moderately high third-order nonlinear susceptibility $\chi^{(3)}$ \cite{moss2013new}. Third-order nonlinear optical effects and applications have been reported in SiN photonic structures, such as supercontinuum generation \cite{johnson2015octave}, frequency comb generation \cite{shen2020integrated}, third harmonic generation \cite{Ning:13}, optical-frequency synthesizer \cite{spencer2018optical}, and so on. Unfortunately, there is no second-order nonlinearity (with a susceptibility of $\chi^{(2)}$) in SiN due to its centrosymmetry. It overshadows the versatility of SiN PIC considering second-order nonlinear effects are important for high-efficiency second harmonic generation, sum-frequency generation, and optical parametric oscillator and amplifier. Though the modification of material composition could induce $\chi^{(2)}$ in SiN \cite{koskinen2017enhancement,ning2012strong}, the realized $\chi^{(2)}$ can not be comparable with that in the materials with intrinsic second-order nonlinearity, which limits the efficiency of second-order nonlinear processes. 

In this work, we propose and demonstrate that by integrating a two-dimensional (2D) GaSe flake on a SiN photonic structure, second-order nonlinear effects could be realized with high-efficiency. While with the recent development of thin-film lithium niobate, it is possible to realize second-order nonlinear processes in all lithium niobate PICs, its fabrication is more difficult and complicated than that of the SiN photonic platform \cite{chen2009photonic,churaev2021heterogeneously}. The realization of lithium niobate thin film in a large wafer and high-quality lithium niobate photonic nanostructure still have challenges. Hetero-integrating a second-order nonlinear materials (such as gallium arsenide and lithium niobate) on SiN photonic platform is expected to have advantages of easy fabrication and high-quality photonic devices to endow the desirable second-order nonlinear effects. In these hetero-integrations, to employ the advantage of easy fabrication in the SiN photonics, these materials are expected to be post-integrated on SiN photonic structures directly without the requirement of further fabrication. Note that SiN has a lower refractive index ($n$ = 1.98) than those materials (the refractive indices in gallium arsenide and lithium niobate are $n$ = 3.37 and $n$ = 2.21, respectively). If the integrated material has a bulk form on the SiN photonic structure, the light will leak from the SiN photonic structures into the covered bulk material, breaking the photonic circuit. Hence, to ensure the confined mode propagating in SiN photonic structures, the integrated materials should be in a form of thin slab with a thickness of tens of nanometers. In this case, the heterostructure of the SiN photonic structure and the covered thin slab could still form a confined waveguiding mode in the circuit. It poses challenges to integrate bulk materials with $\chi^{(2)}$, such as gallium arsenide and lithium niobate, which requires complicated and costly techniques of wafer bonding, ion-implantation-cutting, and chemical mechanical polishing to obtain a thin slab on the SiN PIC \cite{churaev2021heterogeneously,park2020heterogeneous}. 

On the contrary, for 2D materials, it is easy to obtain their slabs with a thickness of tens of nanometers or even few atomic layers by virtue of their layered structure with the weak interlayer van der Waals interactions. This simplifies the integration of few-layer 2D materials on SiN PIC by a direct transfer-printing technique \cite{castellanos2014deterministic}. In addition, since the dangling-bond-free of 2D materials, it is possible to epitaxially grow them on SiN PIC directly\cite{liu2022silicon}. These advantages of 2D materials circumvent complicated and costly processes of bulky $\chi^{(2)}$ material integration with SiN structures. Thanks to the well-developed fabrication processes and excellent performances of 2D materials, the integration of 2D materials on photonic structures offers the opportunity of better device performances and novel features. In nonlinear optical field, four-wave mixing \cite{qu2020enhanced}, Kerr frequency comb \cite{castro2017plasmonically}, and Raman laser \cite{shen2020raman} are demonstrated based on atomically thin materials integrated photonic structures. Remarkably, extremely high $\chi^{(2)}$ of some 2D materials promise strong second-order nonlinear processes in 2D materials integrated photonic structures. For example, $\chi^{(2)}$ of 2D GaSe (2.4 nm/V) \cite{zhou2015strong} is two orders of magnitude higher than the mostly employed bulk $\chi^{(2)}$ material lithium niobate (0.027 nm/V) \cite{miller1971dependence}. With the stacking sequence of $\varepsilon$-GaSe, few-layered GaSe ﬂakes are always noncentrosymmetric. Thus, SHGs are allowed in an arbitrary number of layers \cite{jie2015layer} and would be additive with the layer numbers.

Here, with a few-layer GaSe coated on a SiN microring resonator, strong second-order nonlinear effects including second-harmonic generation (SHG) and sum-frequency generation (SFG) are obtained with the pump of microwatts continuous-wave (CW) lasers. A phase-matching condition between the fundamental pump laser and SHG/SFG signals are considered by engineering their corresponding guiding modes in the microring. The conversion efficiencies of SHG and SFG are respectively estimated as 849\%/W and 123\%/W. Our results indicate that the assistance of few-layer GaSe could be considered as a reliable method to explore second-order nonlinear processes in SiN photonics platform, including optical parametric oscillators (OPOs) \cite{okawachi2015dual}, spontaneous parametric downconversion \cite{Christophe2018}, etc. It also paves the way for on-chip applications of SHG and SFG based on the versatile SiN photonics platform.

\section{Methods}

\subsection{Device design}

\begin{figure}[htbp]
\centering
\includegraphics[width=0.75\textwidth]{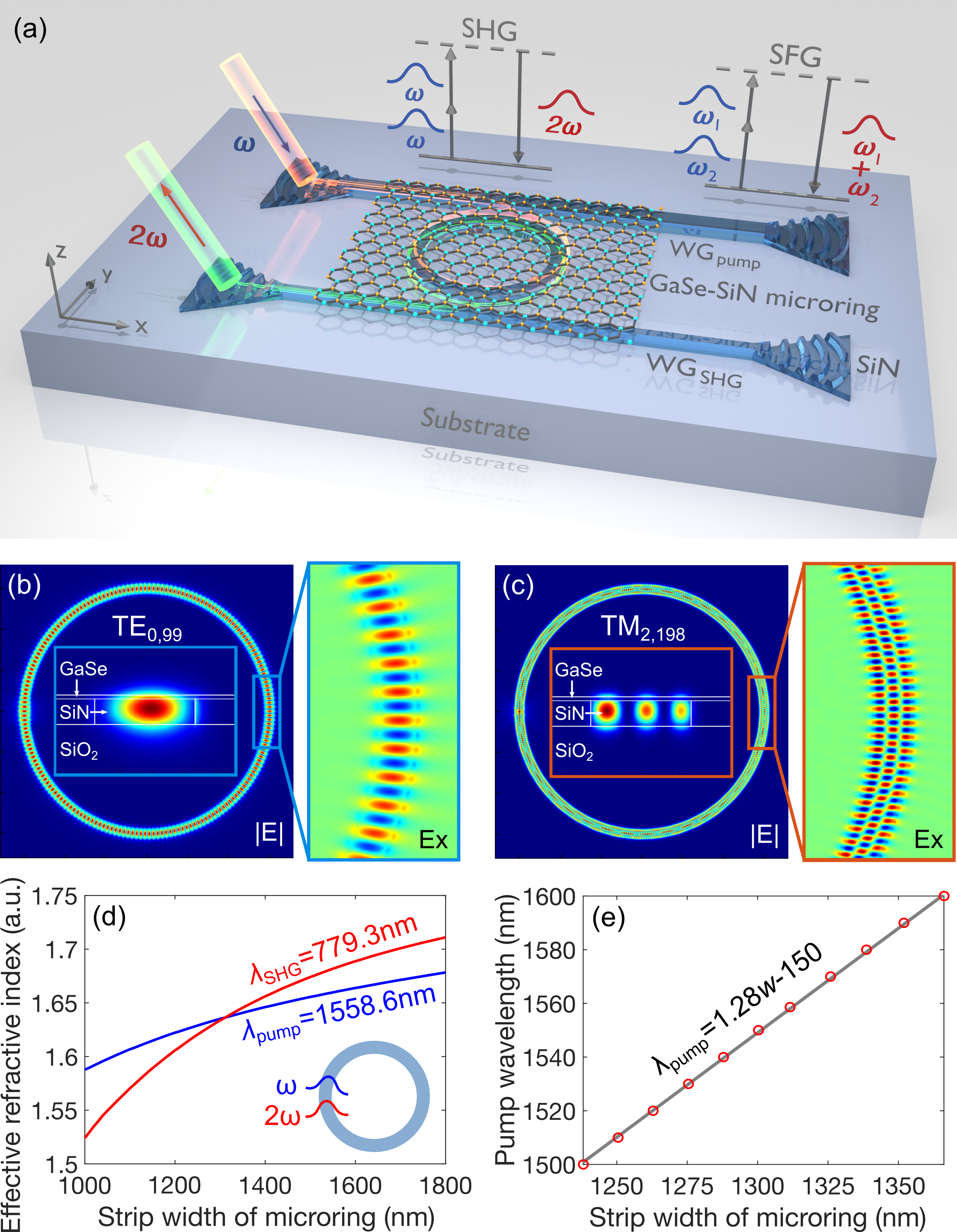}
\caption{(a) Schematic of SHG and SFG processes in the proposed GaSe-SiN microring. Insets are energy diagrams of SHG and SFG. 
(b,c) Mode profiles of fundamental pump mode and SHG mode in the GaSe-SiN microring with the strip width of 1.31 $\upmu$m, respectively. The circular mode profiles are taken from the middle plane of the microring, which are zoomed in the right images. The center insets are cross-sectional distributions of the mode profiles. (d) Dependences of the effective refractive indices of resonance modes at the wavelength of  $\lambda_{\rm pump} = 1558.6$ nm and $\lambda_{\rm SHG} = 779.3$ nm on the strip width of the microring, respectively. 
(e) Optimized strip widths of the mciroring for phase-matching condition of SHG processes pumped at varied wavelengths.}
\label{Fig_1}
\end{figure}

The implementation of second-order nonlinear processes (SHG and SFG) from the GaSe-integrated SiN microring resonator is schematically illustrated in Figure \ref{Fig_1}a. The microring coated with a few-layer GaSe flake is coupled by two bus-waveguides, which are designed for coupling in the fundamental pump laser (${\rm WG_{pump}}$) and coupling out the SHG/SFG signals (${\rm WG_{SHG}}$), respectively. The SiN microring and waveguides are fabricated in a 300 nm thick SiN slab on a 3 $\upmu$m thick SiO$_2$. The integrated few-layer GaSe flake interacts with the evanescent field of the SiN microring, as indicated by the guiding mode distributions of the GaSe-SiN microring shown in the center insets of Figures \ref{Fig_1}b and \ref{Fig_1}c. 

Benefiting from the high quality ($Q$) factors of the resonance modes in the SiN microring, the on-resonance light would circulate in the microring for a long time, which could ensure the effective GaSe-light interaction. As a result, strongly enhanced second-order nonlinear processes are guaranteed in the GaSe flake with the consideration of GaSe’s high $\chi^{(2)}$, and therefore form new guiding modes in the microring at the short wavelength region in the phase matching condition between the fundamental pump laser and second-order nonlinear signals. In the SHG process, for a fundamental pump laser with a frequency of $\omega$ 
incident into the input waveugide ${\rm WG_{pump}}$, it would evanescently couple into the GaSe-SiN microring to form resonance modes. The yielded SHG signal with a frequency of $2\omega$ in the microring would then couple into the output waveguide ${\rm WG_{SHG}}$. Similarly, in the SFG process, with the excitation of two pump lasers at the frequencies of $\omega_1$ and $\omega_2$, SFG signal at the frequency of $\omega_1+\omega_2$ would be finally collected at the output waveguide ${\rm WG_{SHG}}$. The photon conversion processes with virtual energy levels are schematically shown in the insets of Figure \ref{Fig_1}a. 

Note, due to the material and mode dispersions between the fundamental pump laser and frequency upconversion signals, the problem of phase-mismatching between their guiding modes in the microring should be addressed carefully to ensure the SHG/SFG signals generated at different locations along the microring interfere constructively and yield maximum output powers \cite{pernice2012second}. To solve that, we design the SiN microring specially to make the guiding modes of the fundamental pump laser and frequency upconversion signals have the same effective refractive indices. As an example, the radius of the SiN microring is chosen as 15 $\upmu$m, and the fundamental pump laser (SHG signal) has a wavelength of 1558.6 nm (779.3 nm). With the thicknesses of the SiN slab and the coated few-layer GaSe fixed as 300 nm and 40 nm, respectively, the effective refractive indices of guiding modes at 1558.6 nm and 779.3 nm could be engineered by changing the strip width $w$ of the SiN microring. The guiding modes and the corresponding effective refractive indices are calculated using a 2D Finite-Difference Eigenmode solver by considering the cross-section of the GaSe-coated SiN strip waveguide. The refractive index of 2D GaSe is obtained from sellmeier and thermo-optic dispersion formulas of GaSe \cite{kato2013sellmeier}. Due to the small radius of the SiN microring, the waveguide bending is considered in calculating the guiding modes. The TE-type mode (with the polarization parallel to the waveguide plane) is considered for the fundamental pump light due to its low transmission loss, specially in the bending with a small radius. In addition, because the largest second-order susceptibility element of the employed 2D GaSe is $\chi^{(2)}_{33}$ \cite{zhou2015strong}, to realize the maximum frequency conversion efficiency, the TM-type mode of the microring (with the polarization normal to the waveguide plane) should be chosen for the SHG signal when the fundamental pump light is TE-type mode of the microring, as indicated in the center inset of Figure \ref{Fig_1}c.

Figure \ref{Fig_1}d plots the dependences of the effective refractive indices on the strip width of the microring for the fundamental mode and SHG mode. At $w = 1.31$ $\upmu$m, there is a cross point between them, indicating their same effective refractive indices. It implies phase-matching condition is satisfied for the SHG process pumped by 1558.6 nm light at this strip width \cite{chuang2012physics,jiang2020high}. Figures 1b and 1c display the mode profiles for the guiding modes of the fundamental pump laser and the SHG signal, which are respectively identified as TE$_{0,99}$ mode and TM$_{2,198}$ mode. The two subscripts are radial and azimuthal order numbers, respectively. The radial order number $n$ of fundamental pump mode and SHG mode are identified by the radial mode profiles, as shown in the central insets of Figure \ref{Fig_1}b and \ref{Fig_1}c. The azimuthal order numbers $m$ are identified by the theoretical expression of microring’s resonance wavelength \cite{van2016optical} $\lambda_m=2\pi R n_{\rm eff}/m$  combining with the simulated $n_{\rm eff}$. $R$ is the radius of the microring. For example, with $n_{\rm eff} = 1.64$ obtained in Figure \ref{Fig_1}d, the results are $m = 99$ for the fundamental pump mode and $m = 198$ for the SHG mode. According to the phase matching condition $k_{\rm SHG}=2k_{\rm pump}$ \cite{kuo20094} and the relationship between the wavevector $k$ and the azimuthal order number $m$ ($k_{\rm pump,SHG}=m_{\rm pump,SHG}/R$) \cite{yang2007enhanced}, the relationship of $m$ between the fundamental pump mode and the SHG mode clearly proves the phase-matching in the SHG process.

With the same analysis method, we also calculate the optimized strip widths of microring for satisfying phase-matching condition in the SHG processes pumped at other wavelengths. The result is shown in Figure \ref{Fig_1}e. The optimized strip widths $w$ and the pump wavelengths $\lambda_{\rm pump}$ have a relation of $\lambda_{\rm pump}=1.28w-150$. Hence, it is straightforward to design the GaSe-SiN microring for maximizing the SHG process at desirable wavelengths. For the SFG process, the phase-matching condition could be determined with the similar method. By making the sum of effective refractive indices of two different fundamental pump lasers equals the effective refractive index of the SFG signal, phase compensation could be achieved for the guiding modes of the three waves and then phase-matching condition is guaranteed.

To maximize the coupling-in (coupling-out) efficiency of the fundamental pump laser (SHG signal) from the $\rm WG_{pump}$ (microring) to the microring ($\rm WG_{SHG}$), the two bus waveguides $\rm WG_{pump}$ and $\rm WG_{SHG}$ are designed with different widths, as indicated in Figure \ref{Fig_1}a. Their widths are optimized by the effective refractive index engineering. The width of $\rm WG_{pump}$ was selected by matching $n_{\rm eff}$ of fundamental TE mode of straight SiN waveguide to $n_{\rm eff}$ of fundamental TE mode of microring. The width of $\rm WG_{SHG}$ was optimized to make the $n_{\rm eff}$ of second order TM mode of microring and the fundamental TM mode of waveguide equal. The strip widths of the $\rm WG_{pump}$ and $\rm WG_{SHG}$ are optimized as 1.35 $\upmu$m and 0.32 $\upmu$m, which respectively propagate the fundamental TE and fundamental TM modes (Insets in Figure \ref{Fig_2}b). In a such narrow $\rm WG_{SHG}$, the guiding mode of the fundamental pump laser in the microring is filtered, hence only SHG signal is collected from $\rm WG_{SHG}$. The coupling-gaps between the two waveguides and the microring are designed according to the critical coupling condition. To maximize the SHG and SFG conversion efficiencies, two conditions are required: (1) The coupling efficiency between the bus-waveguides and the microring should be as high as possible to ensure enough pump light couple into the microring and enough SHG signal couple out from the microring; (2) The $Q$ factor of the microring should be as high as possible to ensure the optical field in the resonant mode is enhanced effectively. To achieve them, the critical coupling between the $\rm WG_{pump}$ ($\rm WG_{SHG}$) and the microring should be satisfied by controlling their coupling gap \cite{guo2016second}, which is optimized as 60 nm (30 nm). At the two ends of each waveguide, grating couplers are designed to assist the light coupling between the external single-mode fibers and the waveguides. The grating couplers are designed according to the Bragg condition by taking a duty cycle of 0.5 and their first diffraction order has a small angle $\theta = 15^\circ$ to the normal of the grating surface.

\subsection{Device fabrication}

The fabrication of the proposed GaSe-SiN microring was started from a 300 nm thick SiN slab grown on a 3 $\upmu$m thick SiO$_2$ buried layer on a Si substrate. The SiN microring resonator and bus waveguides were defined by electron beam lithography. Following by an inductively coupled plasma dry etching of SiN, the residual e-beam resist was removed by the  piranha solution. A few-layer GaSe flake was prepared using mechanical exfoliation method from a monocrystalline bulky GaSe grown by Bridgeman-Stockbarge method, which was then transferred on top of the SiN microring using a dry transfer method \cite{gan2018microwatts}. 

\subsection{Optical characterisations}

The optical characterization of the fabricated device was implemented using a vertical grating coupling alignment system. A narrowband tunable CW laser in the telecom-band is chosen as the fundamental pump laser, which is guided in a single-mode fiber and coupled into the grating coupler of the pump waveguide $\rm WG_{pump}$. Its polarization is controlled using a fiber-based polarization controller to excite the TE guiding mode in the SiN waveguide and microring. By fixing the power of the CW laser and tuning its wavelength from 1500 nm to 1600 nm with a step of 0.1 nm, the transmission spectrum from the $\rm WG_{pump}$ is obtained by measuring its transmission power at the output port.

\section{Results}
\subsection{Characterization of GaSe-SiN microring resonator}

\begin{figure}[ht]
\centering
\includegraphics[width=0.65\textwidth]{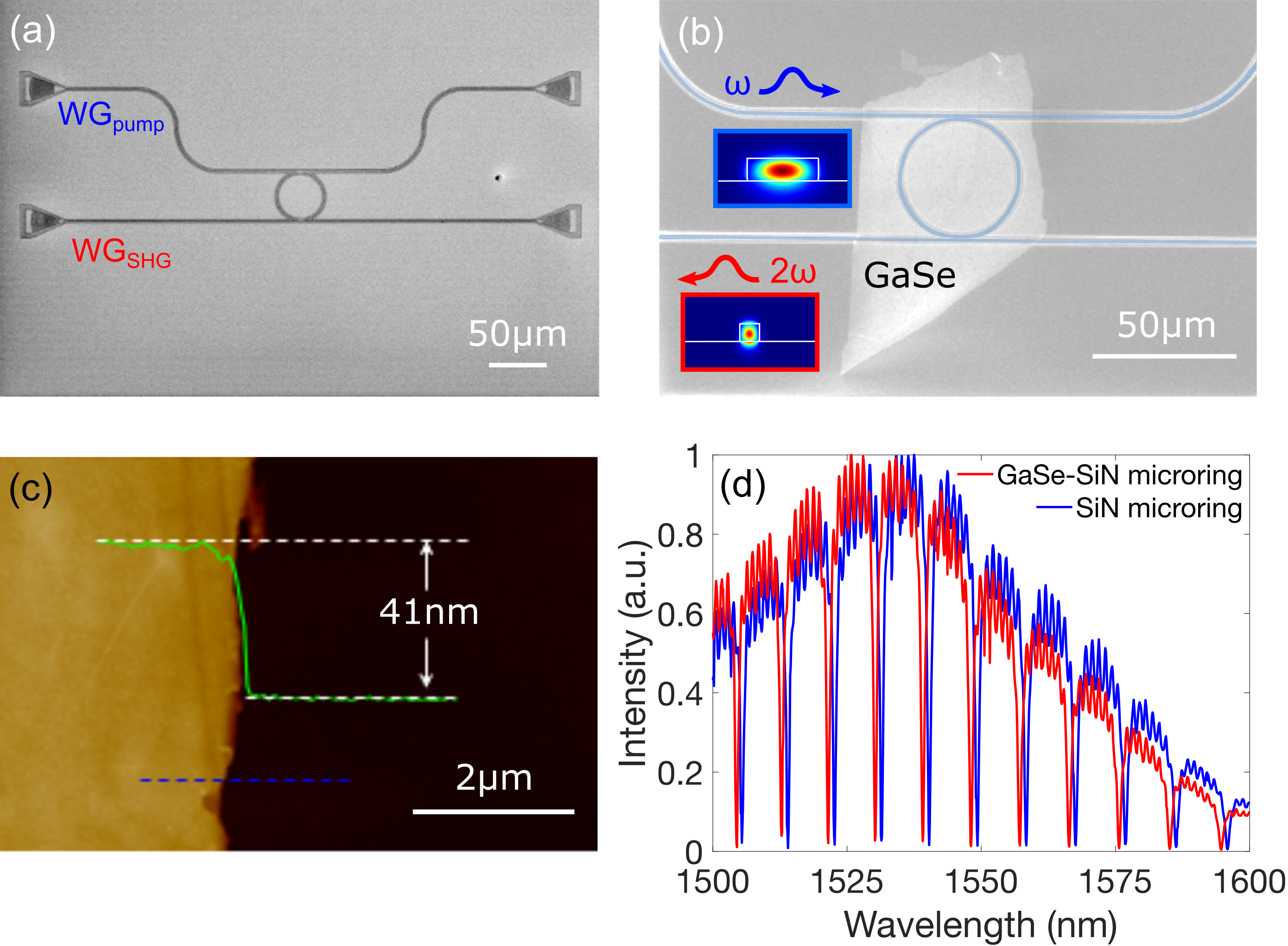}
\caption{(a) SEM image of the fabricated SiN microring coupled with two bus-waveguides. (b) SEM image of the finished GaSe-SiN microring, showing the GaSe flake tightly contacting on the microring. Insets are the cross-sections of the simulated guiding modes in the $\rm WG_{pump}$ and $\rm WG_{SHG}$. (c) AFM image of the employed few-layer GaSe flake in the GaSe-SiN microring. The inset indicates the thickness of GaSe measured along the blue dashed line. (d) Normalized transmission spectra of the SiN microring measured via the input and output ports of the $\rm WG_{pump}$. Blue and red curves are results obtained without and with the transferred GaSe flake.}
\label{Fig_2}
\end{figure}

Figures \ref{Fig_2}a and \ref{Fig_2}b display scanning electron microscope (SEM) images of the fabricated SiN microring resonator before and after the integration of the few-layer GaSe flake, respectively. The few-layer GaSe flake covers the whole resonator region, promising effective interaction between GaSe flake and evanescent field of the resonance modes. The thickness of the transferred GaSe flake is examined as 41 nm using the atomic force microscope (Figure \ref{Fig_2}c), corresponding to a layer number of 48 by assuming the thickness of monolayer GaSe as 0.85 nm \cite{tang2016layer}. The morphology of the device shown in the SEM image indicates the tight contact of the GaSe ﬂake on the SiN microring.

The pump laser is a narrowband tunable CW laser. The transmission spectra of the SiN microring before and after the integration of  GaSe are displayed in Figure \ref{Fig_2}d. Periodic transmission dips with a free spectral range of 8.7 nm are observed, which indicate the microring’s resonance modes coupled with the guiding mode of the $\rm WG_{pump}$. In addition, the transmission dips are close to zero power, indicating the critical coupling between the $\rm WG_{pump}$ and the microring. The unflat transmission baseline of the resonance dips is arisen from the limited coupling bandwidth of the grating couplers. Comparing the transmission spectra with and without the GaSe flake, the resonance dips red-shift by 1.5 nm, which could be attributed to GaSe’s high refractive index perturbating the resonance condition of the microring. In addition, by fitting the resonance dips with Lorentizan lineshapes, the $Q$ factors of the resonance modes without and with GaSe  are evaluated as 2000 and 1800, respectively. Compared with the previously reported SiN microring, the $Q$ factors obtained in our device is much lower. It could be attributed to the large mode scattering loss due to the strong overlap between the optical mode and the defected boundaries of the SiN microring \cite{bauters2011ultra}. In previous work, the SiN microrings were fabricated with extremely thin films (smaller than 100 nm) covered with a thick SiO$_2$ cladding layer or with much thicker films (around 1000 nm) \cite{ji2021methods}, which both have large cross-sectional mode profiles to reduce the scattering influence by the boundary defects. In our devices, the $Q$ factors are expected to be improved greatly by depositing the SiO$_2$ cladding layer to expand the mode profile.

\subsection{Second-harmonic generation (SHG)}

\begin{figure}[ht]
\centering
\includegraphics[width=0.75\textwidth]{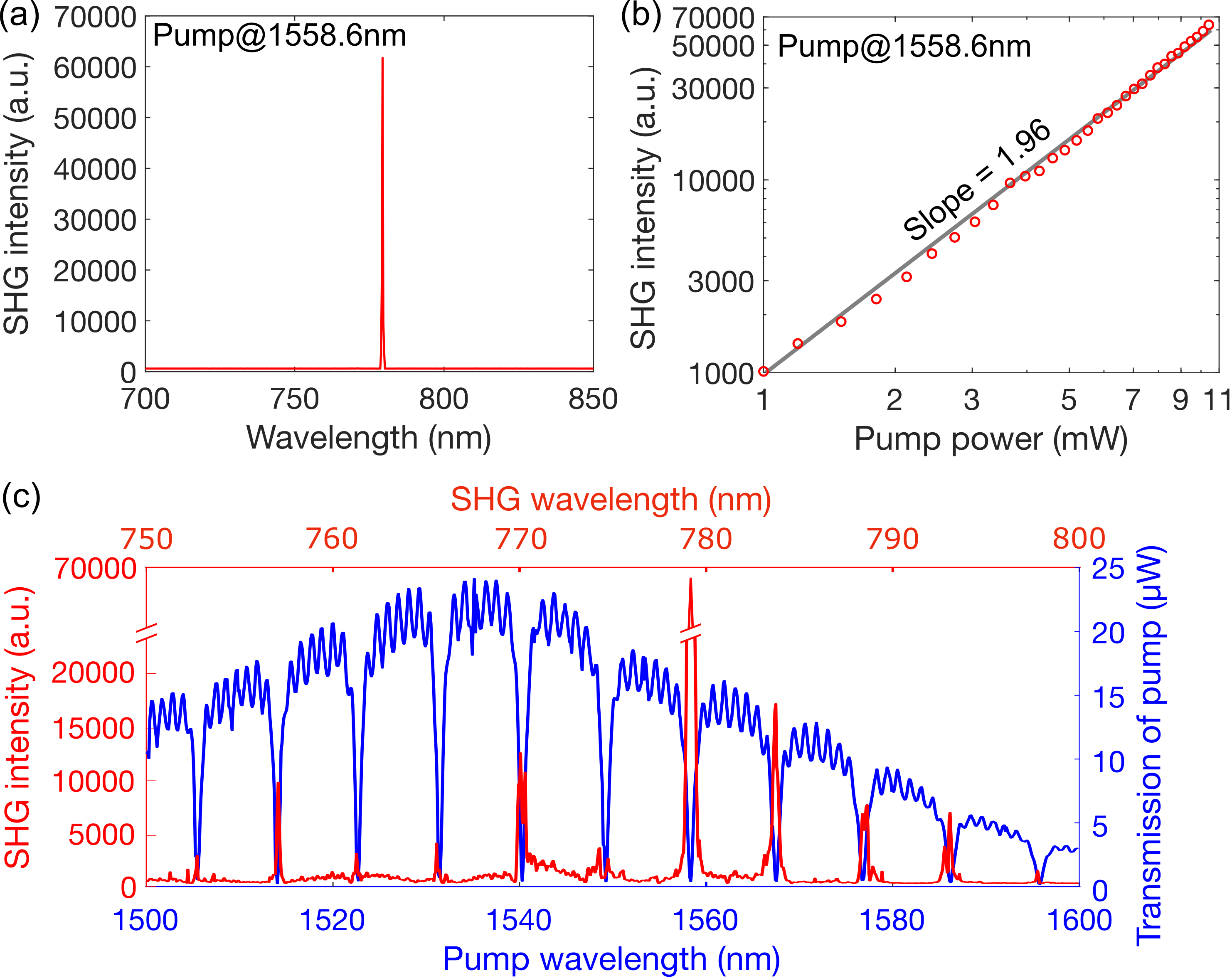}
\caption{SHG in the fabricated GaSe-SiN microring. (a) SHG spectrum measured with a CW pump laser at the wavelength of 1558.6 nm, which is on-resonance with the GaSe-SiN microring. (b) Dependence of the SHG intensity on the pump power at the wavelength of 1558.6 nm in the log-log scale, where the black line is a linear fitting with a slope of 1.96. (c) Transmission spectra of the fundamental pump light and the varied SHG intensities measured from  $\rm WG_{pump}$ and $\rm WG_{SHG}$, respectively, when the CW pump laser is tuned from 1500 nm to 1600 nm.}
\label{Fig_3}
\end{figure}

The resonance modes in the microring resonator could enhance GaSe-light interaction through its evanescent field, which promises the effective SHG/SFG processes. By tuning the CW laser wavelength as 1558.6 nm to be on-resonance with a resonance mode of the GaSe-SiN microring, SHG signal is examined from the $\rm WG_{SHG}$. As shown in Figure \ref{Fig_3}a, a strong frequency upconversion signal is obtained at the wavelength of 779.3 nm. The SHG nature of this signal is identified by two times ratio between the wavelengths of fundamental pump light and SHG signal, which is then confirmed by the pump power dependent SHG intensity. With the gradually varied pump powers, the SHG intensities are acquired, as shown in the log-log plot of Figure \ref{Fig_3}b. The fitted line indicates a slope of 1.96, verifying the quadratic power dependence of signal, which is the typical characteristic of SHG. 

To further explore the origin of the strong SHG signal, we then examine the dependence of the SHG intensity on the pump wavelength. By scanning the pump laser from 1500 nm to 1600 nm with a step of 0.1 nm, the corresponding SHG intensities are measured from the $\rm WG_{SHG}$, as shown in the red curve of Figure \ref{Fig_3}c. For comparison, in Figure \ref{Fig_3}c, the transmission spectrum of the pump laser measured from the output port of $\rm WG_{pump}$ is plotted as well. When the laser wavelength locates at the resonance dips of the transmission spectrum, which satisfies the condition of on-resonance pump, the SHG signal presents a strong intensity. For the pump laser with wavelengths off-resonance from the microring, there is no observable SHG signal. Note, in the bare SiN microring, we also implement the SHG measurements by aligning the pump laser on-resonance. No SHG signal is observed either. This clearly proves that the achieved SHG signals originate from the assistance of the integrated GaSe layer with strong second-order nonlinear susceptibility as well as the resonance-enhanced GaSe-light interaction in the GaSe-SiN microring. For the on-resonance pump at the wavelength of 1558.6 nm, the SHG signal is much stronger than those obtained at other on-resonance pump wavelengths. It could be attributed to the satisfied phase-matching condition between the pump laser  (at 1558.6 nm) and SHG signal (at 779.3 nm), which is the case discussed in Figure \ref{Fig_1}. 

We then evaluate the absolute conversion efficiency of the enhanced SHG from the GaSe-SiN microring. To do that, the optical losses resulting from different parts of the light path are estimated first. 
To evaluate the coupling efficiency of the grating couplers, a laser at an off-resonance wavelength of 1556.0 nm is coupled into the $\rm WG_{pump}$, which could be considered without coupling with the microring. With the laser power of 10 mW, the transmission power of the $\rm WG_{pump}$ is about 14 $\upmu$W (Figure \ref{Fig_3}c). Considering the scattering loss of 0.78 dB due to 41 nm thick GaSe flake, and negligible propagation loss of the straight SiN waveguide, the coupling loss of every grating coupler is estimated through the $\rm WG_{pump}$ light path, which is about 13.62 dB. Based on these calculations, the fundamental pump laser at 1558.6 nm coupled into GaSe-SiN microring is estimated as $P_{\rm pump}=355$ $\upmu$W. With this on-resonance pump, the SHG signal coupled from the $\rm WG_{SHG}$ is calibrated with a photomultiplier tube, showing an estimated power of $P_{\rm SHG}=1.07$ $\upmu$W. Therefore, the conversion efficiency of SHG is calculated as $\eta=P_{\rm SHG}/P_{\rm pump}^2 \times 100\%=849$\%/W \cite{guo2016second}.
Moreover, we observed strong SHG signal even when the pump power is even low as 35 $\upmu$W. This optical power is easily reachable in chip-integrated III-V semiconductor lasers \cite{liang2021recent}, indicating the proposed GaSe-SiN microring indeed has the potential to carry out on-chip SHG. There are some other reported works about SHG in SiN photonic structures. For example, with a SiO$_2$ cladding on the SiN microring, their interface could support a $\chi^{(2)}$ due to the surface broken inversion symmetry, however, the enabled SHG conversion efficiency is low as 0.1\%/W even with the resonance-enhancement \cite{levy2011harmonic}. Another approach combining $\chi^{(3)}$ of SiN with an photogalvanic effect induced direct-current electric field could also be employed to realize SHG and the conversion efficiency exceeds 2500\%/W \cite{lu2021efficient}. Unfortunately, the slow photogalvanic effect degenerates the SHG response time from femtoseconds to minutes. In addition, lithium niobate on insulator (LNOI) platform is reported with high SHG conversion efficiency \cite{chen2019ultra}. However, the CMOS-incompatible Ar$^+$ plasma dry etching of lithium niobate increases its fabrication complexity and cost and hence is restricting its commercialization applications \cite{zhu2021integrated}. By contrast, the possibility to directly epitaxial grow 2D GaSe on SiN structures \cite{liu2022silicon} promises much simple and low-cost SHG strategy.


\subsection{Sum frequency generation (SFG)}

\begin{figure}[ht]
\centering
\includegraphics[width=0.75\textwidth]{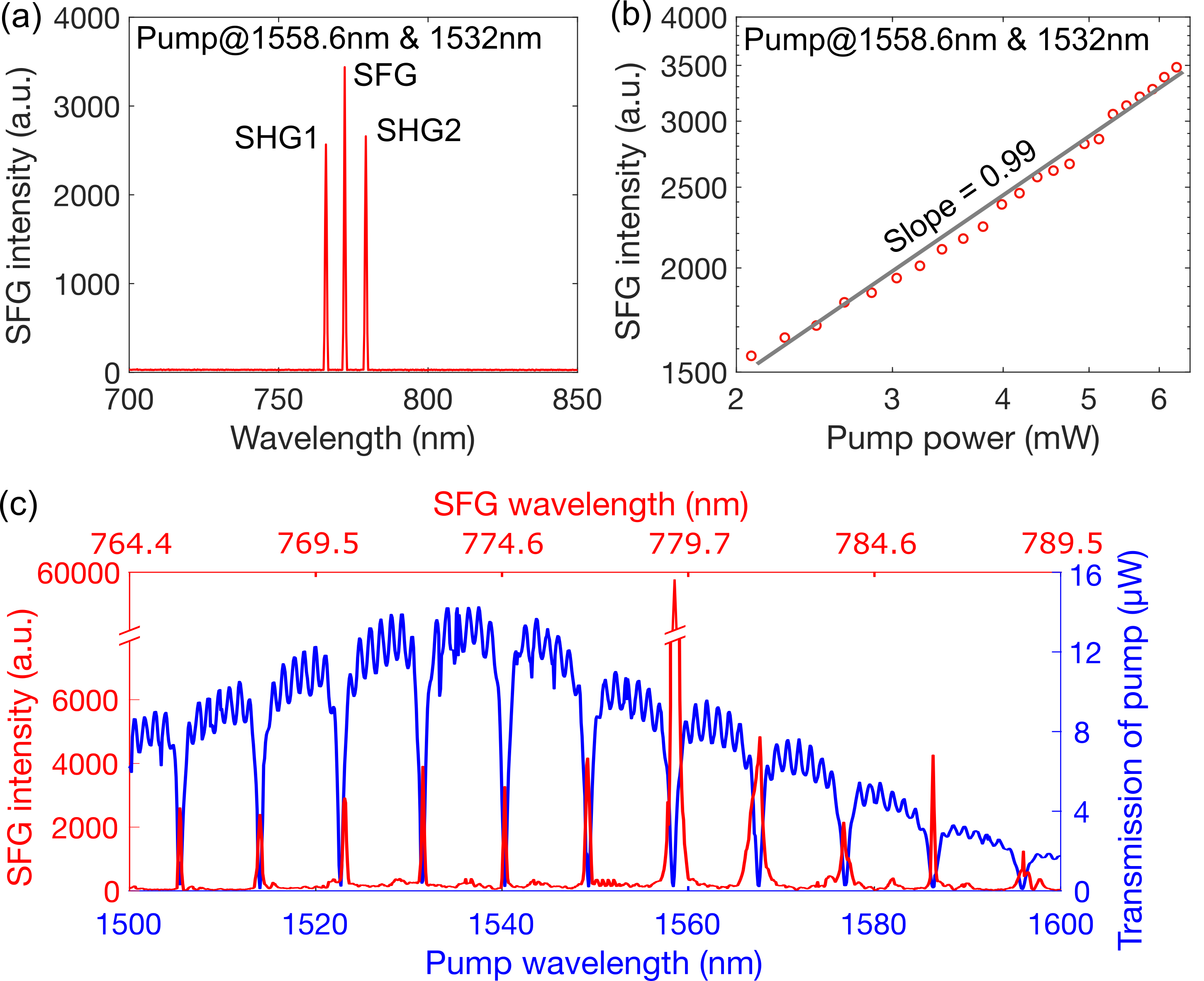}
\caption{SFG in the fabricated GaSe-SiN microring. (a) Spectrum of the frequency upconversion signals measured from the $\rm WG_{SHG}$ of the device under the pump of two CW lasers at 1558.6 nm and 1532.0 nm. (b) SFG intensities measured with the varied power of Laser2, while the power of Laser1 is fixed. (c) SFG intensities measured with the wavelength of Laser1 fixed at 1558.6 nm, and the wavelength of Laser2 scanned from 1500 nm to 1600 nm, where the transmission spectrum of the fundamental pump lasers is also shown for comparison.}
\label{Fig_4}
\end{figure}

Another important second-order nonlinear parametric process is the SFG, which involves two different pump lasers and generates photons with a frequency of their summation. The above  CW pumped high-efficiency SHG from the GaSe-SiN microring indicates the possible SFG by considering the resonance-enhancement and GaSe’s high $\chi^{(2)}$. To carry out that, we employ two CW tunable lasers and couple them into the $\rm WG_{pump}$ together to excite the modes in the GaSe-SiN microring. One of the lasers (named as Laser1) is tuned to the wavelength of 1558.6 nm, which is on-resonance with the microring. The wavelength of another laser (named as Laser2) could be tuned over a range from 1500 nm to 1600 nm. First, by tuning the wavelength of Laser2 as 1532.0 nm, the spectrum of the frequency upconversion signals from the GaSe-SiN microring is measured from the output port of $\rm WG_{SHG}$, as shown in Figure \ref{Fig_4}a. Three signal peaks are obtained at the wavelengths of 766.0 nm, 772.6 nm, and 779.3 nm. According to the wavelength conversion relationship, the first ($\lambda$ = 766.0 nm) and the third ($\lambda$ = 779.3 nm) peaks are recognized as SHG signals excited by the CW pump Laser2 and Laser1, respectively. The second peak ($\lambda$ = 772.6 nm) is identified as SFG signal excited by the two CW pump lasers. This is confirmed by the relationship between the pump power and the SFG intensity. We lock the pump power at 6 mW of Laser1 and gradually increase the pump power of Laser2. As shown in Figure \ref{Fig_4}b, a slope of 0.99 is obtained by a linear fitting, which is close to the theoretical value 1 of typical log-log curve for SFG process. 

To carefully study the resonance enhanced SFG, we further measure the pump wavelength dependent SFG signals, as shown in Figure \ref{Fig_4}c. The wavelength of Laser1 is fixed at 1558.6 nm, and the wavelength of Laser2 is scanned. In Figure \ref{Fig_4}c, the transmission spectrum of the GaSe-SiN microring is also plotted for comparison. The acquired SFG signals show strong peaks when the wavelength of Laser2 locates on the resonance wavelengths of the microring. If Laser2 is off-resonance from the microring, there is no SFG signal. Considering Laser1 is fixed on-resonance, it is concluded that the achievement of SFG requires both pump lasers to be on-resonance. In Figure \ref{Fig_4}c, the highest SFG peak is at 779.3 nm, which is actually obtained when both of the two pump lasers are tuned at 1558.6 nm. As discussed in Figures 1 and 3, the GaSe-SiN microring is designed for this pump laser at 1558.6 nm to satisfy the phase-matching condition, which therefore show much stronger frequency upconversion signal than those obtained with other pump lasers.

Similar as the calculation of SHG, the conversion efficiency of SFG is expressed as $\eta=P_{\rm SFG}/P_{\rm Laser1}P_{\rm Laser2} \times 100\%$, where $P_{\rm SFG}$, $P_{\rm Laser1}$, and $P_{\rm Laser2}$ are the excited SFG power and pump powers of Laser1 and Laser2 coupled into the GaSe-SiN  microring. For the case that Laser2 has a wavelength of 1532.0 nm, the measured SFG power at the wavelength of 772.6 nm is calibrated as 2.05 nW, corresponding to $P_{\rm SFG}=57.78$ nW excited in GaSe-SiN microring. Given the pump powers of both Laser1 and Laser2 as $P_{\rm Laser1}=P_{\rm Laser2}=212.88$ $\upmu$W, the calculated conversion efficiency of SFG is 123\%/W. Considering the high conversion efficiency of SFG and high controllable capability for the new frequency generation, the GaSe-SiN microring is promising in the on-chip SFG applications, such as low-frequency weak light detection \cite{strekalov2009efficient} and single photon up-conversion detectors/spectrometer \cite{MA201269}.

\section{Conclusion}

In conclusion, we have demonstrated high-efficiency second-order nonlinear parametric processes (SHG and SFG) in a SiN microring with the assistance of an integrated few-layer GaSe flake, though they are forbidden in the centrosymmetric SiN. Thanks to the strongly enhanced GaSe-light interaction by the microring’s resonance modes as well as GaSe’s high $\chi^{(2)}$, the second-order nonlinear processes could take place effectively. Further, we also considered the phase-matching condition by optimizing the microring structure and engineering the guiding modes to maximize the final output SHG and SFG signals. High conversion efficiencies of 849\%/W and 123\%/W are respectively obtained for SHG and SFG with low-power CW pump lasers, which could be lowered to the level of microwatts. Considering SiN photonic platform has been recognized very promising in a variety of photonic technologies, with the integrated few-layer GaSe to complement the limitation of the absent second-order nonlinearity, the proposed GaSe-SiN photonics platform may pave a new way towards developing high-performance on-chip nonlinear devices based on second order nonlinearity, such as on-chip frequency-conversion laser sources, optical autocorrelators, entanglement photon-pair generations. 



\subsection*{Funding Sources}
This work was supported by the Key Research and Development Program (2018YFA0307200, 2017YFA0303800), the National Natural Science Foundation (61775183, 91950119, 61905196, 11634010, and 62105265), the Key Research and Development Program in Shaanxi Province of China (2020JZ-10), the Fundamental Research Funds for the Central Universities (D5000210905, 310201911cx032 and 3102019JC008).

\subsection*{Notes}
The authors declare no competing financial interest.

\begin{acknowledgement}

The authors thank the Analytical \& Testing Center of NPU for their assistance in device fabrication and characterizations.



\end{acknowledgement}






\bibliography{achemso}

\end{document}